\documentclass[final]{IEEEtran}
\IEEEoverridecommandlockouts
\usepackage{cite}
\usepackage{amsmath,amssymb,amsfonts,amsthm}
\usepackage{algorithmic}
\usepackage{algorithm}
\usepackage{graphicx}
\usepackage{textcomp}
\usepackage{xcolor}
\usepackage{etoolbox}
\usepackage{pdfpages}
\usepackage{url}
\usepackage{lipsum}
\usepackage{multirow}
\usepackage[normalem]{ulem}
\usepackage{bbding}
\usepackage{pifont}
\usepackage{wasysym}
\usepackage{bbm}
\usepackage{dsfont}
\usepackage{mathtools, nccmath}
\usepackage{makecell}
\usepackage{aligned-overset}
\usepackage{stfloats}
\usepackage{booktabs}
\usepackage{array}
\usepackage{threeparttable}
\usepackage{epstopdf}

\DeclareMathAlphabet{\mathbcal}{OMS}{cmsy}{b}{n}
\def\BibTeX{{\rm B\kern-.05em{\sc i\kern-.025em b}\kern-.08em
		T\kern-.1667em\lower.7ex\hbox{E}\kern-.125emX}}


\makeatletter %
\let\myorg@bibitem\bibitem
\def\bibitem#1#2\par{%
	\@ifundefined{bibitem@#1}{%
		\myorg@bibitem{#1}#2\par
	}{%
		\begingroup
		\color{\csname bibitem@#1\endcsname}%
		\myorg@bibitem{#1}#2\par
		\endgroup
	}%
}
\makeatother
\begin{document}
	\title{Bending the Rules of Propagation: Caustic Beamforming for Next-Generation Wireless Systems
		\\%\thanks{\it In case I forget.}
	}
	\author{Shicong~Liu,~\IEEEmembership{Graduate Student Member,~IEEE}, Xianghao~Yu,~\IEEEmembership{Senior Member,~IEEE}, and~Robert~Schober,~\IEEEmembership{Fellow,~IEEE}
		\thanks{
			
			(Corresponding author: Xianghao Yu).
			
			Shicong Liu and Xianghao Yu are with the Department of Electrical Engineering, City University of Hong Kong, Hong Kong (email: sc.liu@my.cityu.edu.hk, alex.yu@cityu.edu.hk).
			
			Robert Schober is with the Institute for Digital Communications, FAU, 91058 Erlangen, Germany (email: robert.schober@fau.de).
			
		}
	}
	
	\maketitle
	
	\begin{abstract}
		Conventional beamforming techniques primarily steer energy along desired directions or focus it at specific locations. These techniques become fragile when facing frequent blockage and highly dynamic propagation environments. 
		In this article, we present caustic beamforming as a new paradigm for wireless beam control. First, we classify representative caustic beams according to their underlying mathematical origins and present three unique properties, namely self-bending, self-healing, and near-field non-diffracting. 
		Building on these propagation properties, we then propose several application scenarios in sixth-generation (6G) networks. We undertake two case studies focused on physical layer security and service stability that highlight the capability of caustic beams to bypass potential eavesdroppers, deliver more uniform coverage, and sustain blockage-resilient links. 
		We further discuss the enabling hardware architectures that facilitate practical deployments, and finally outline key open challenges regarding caustic beams that require further research.
	\end{abstract}

	\section{Introduction}
	\bstctlcite{IEEEexample:BSTcontrol}

The sixth-generation (6G) wireless networks are envisioned to provide transformative capabilities to support integrated sensing and communication (ISAC), embodied and agentic artificial intelligence (AI), and immersive communication~\cite{ITU-R_M.2160-0}. 
These emerging use cases impose stringent requirements on wireless systems in terms of data rate, latency, and service density. 
To address these challenges, current wireless systems rely heavily on beamforming techniques, including far-field beam steering and near-field beam focusing~\cite{10068140}. These prevailing techniques shape the radiated electromagnetic (EM) waves to coherently superpose towards designated far-field directions or at near-field locations.

However, these conventional beamforming techniques face fundamental limitations in complex and dynamic environments. 
First of all, the beamforming gain of conventional schemes is largely restricted to specific far-field directions or near-field focal regions, and degrades rapidly as the user equipment (UE) moves away from them. Consequently, frequent beam training and channel estimation are required to maintain the quality of services of mobile users~\cite{guerboukha_curving_2024}. 
Second, the obstacles envisioned in the complex propagation environments of 6G networks can cause wireless link failures. 
As wireless systems progressively migrate towards the millimeter-wave (mm-wave) and terahertz (THz) frequency bands to support higher bandwidths, the propagation of EM waves suffers from stronger atmospheric absorption and weaker diffraction. As a result, wireless links become dominantly reliant on line-of-sight (LoS) paths, and even a minor obstruction can lead to abrupt link failures~\cite{11267239}.

To address these transmission and coverage challenges, an appealing strategy is to move beyond the conventional beam steering and focusing schemes towards a more resilient beamforming paradigm. 
In recent years, caustic beamforming has emerged as a promising solution to address the fragility of wireless links~\cite{darsena2025airybeamsnearfieldcommunications}. 
In optics, a caustic refers to the envelope of light rays reflected or refracted by a curved surface, creating regions of concentrated intensity~\cite{Zannotti2020}. By leveraging this phenomenon, we can manipulate the spatial superposition of EM waves with unprecedented freedom, enabling novel energy radiating patterns that surpass conventional steering and focusing beams. In particular, 
caustic beams can also be engineered to follow specific curved trajectories to bypass obstructions on LoS paths and to efficiently serve multiple distinct regions simultaneously~\cite{guerboukha_curving_2024,11267239}.

Fortunately, the recent progress in large-scale antenna arrays and reconfigurable EM surfaces enables fine-grained control of phase and amplitude across large apertures. 
Consequently, practical hardware implementations capable of generating caustic beams are becoming increasingly feasible~\cite{10964575}. Moreover, the uplift of carrier frequencies to mm-wave and THz bands allows wavefront engineering methods originating from optics to be seamlessly adapted for wireless communication systems. This convergence of optical theory and wireless practice makes the synthesis of caustic beams a viable reality for future 6G wireless networks.

This article introduces a comprehensive vision for introducing caustic beamforming into next-generation wireless communication networks. 
We begin by unveiling the physics behind caustic waveforms and revealing their unique characteristics. We then propose several application scenarios, where two case studies, i.e., physical layer security and reliable wireless coverage, are undertaken to demonstrate the superiority of caustic beams. 
To support these applications, we further identify practical hardware architectures for caustic-beam synthesis. 
Finally, we discuss related key challenges as well as future research directions.

	\section{Caustics: From Optics to Microwaves} %
	\begin{figure}
		\centering
		\includegraphics[width=0.49\textwidth]{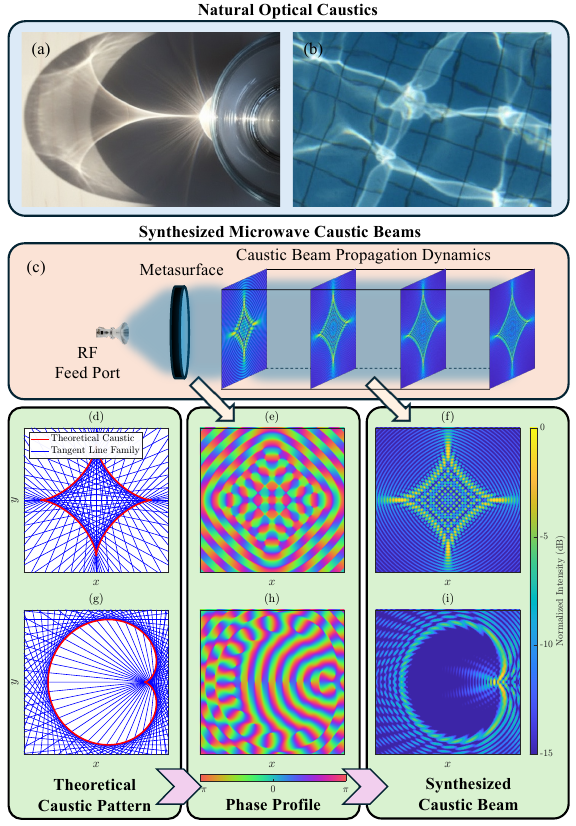}
		\caption{Natural optical caustics produced by refraction from (a) a glass of water and (b) a fluctuating pool surface. (c) Schematic of metasurface-based synthesis of microwave caustic beams. (d) Prescribed astroid caustic pattern and its associated family of tangent lines. (e) Corresponding phase profile of the metasurface. (f) Synthesized astroid caustic beam pattern. (g)-(i) A cardioid example following the same workflow in (d)-(f).
		}
		\label{fig:1}
	\end{figure}

	In optical systems, caustics are often considered natural phenomena resulting from the reflection or refraction of light on curved surfaces. When refracted or reflected light rays become tangent to one another, they undergo coherent superposition, thereby forming bright trajectories or regions in space. 
	A familiar example is the {bright curves formed by sunlight passing through a glass of water, or the shimmering ripples of lights on the bottom of a pool}, as depicted in Figs.~\ref{fig:1}(a) and (b), respectively. 
	
	While typically viewed as optical aberrations in imaging systems, this phenomenon offers a paradigm shift for wireless communications to create beams with arbitrary coverage patterns. 
	Specifically, unlike conventional beamforming techniques, which restrict energy concentration to a specific far-field direction or a near-field focal point, caustic beamforming can aggregate radiated energy along arbitrary spatial trajectories or regions, which significantly improves the flexibility of beamforming.

	The synthesis of microwave caustic beams is summarized in Fig.~\ref{fig:1}(c). Taking the astroid caustic as an example, the family of tangent lines, whose envelope forms the theoretical caustic, is first calculated to determine the local propagation directions of EM waves, as shown in Fig.~\ref{fig:1}(d). The phase profile illustrated in Fig.~\ref{fig:1}(e) is then computed from the tangent family, and is implemented on the metasurface. When radio-frequency EM waves illuminate the metasurface, the imposed phase profile reshapes the outgoing EM waves so that they become tangent to the target curve and form the prescribed astroid shaped caustic beam. The propagation behavior at different distances is also sketched in Fig.~\ref{fig:1}(c), while Fig.~\ref{fig:1}(f) gives a representative transverse structure during propagation. Figs.~\ref{fig:1}(g)-(i) show another example of a cardioid caustic beam synthesized by the same workflow. 
	With proper design, caustic beams can be generated in a variety of transverse shapes to bypass or cover designated regions as required by different wireless communication tasks. 

	\section{Unique Characteristics}\label{sec:iii}
	
	\begin{figure}[t]
		\centering
		\includegraphics[width=0.5\textwidth]{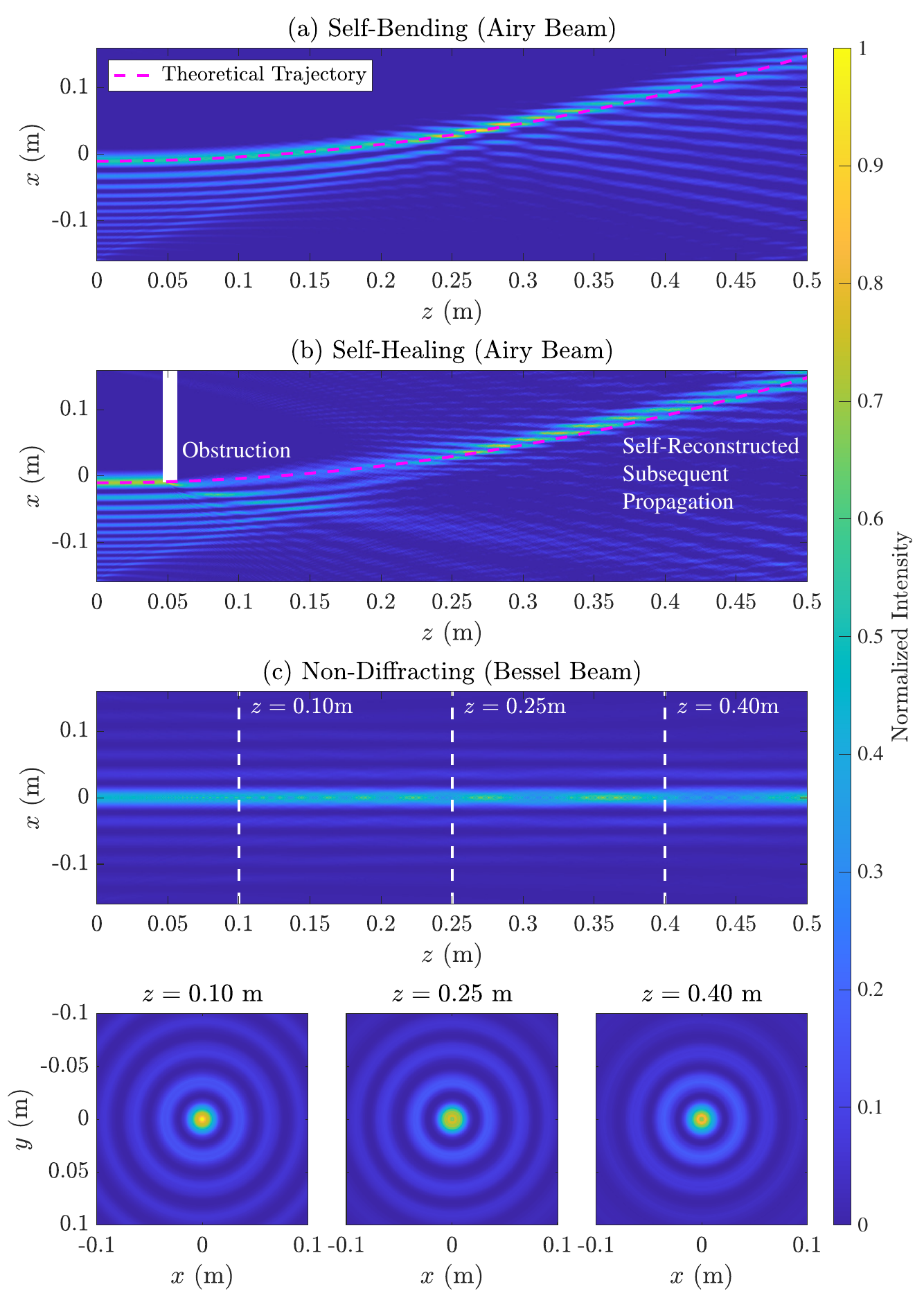}%
		\caption{Visualization of unique properties with an array in the $xOy$ plane and an EM wave propagating along the $z$-axis. (a) Bending trajectory of the Airy beam. (b) Transverse structure recovery after being partially obstructed during propagation. (c) The transverse structure of a Bessel beam remains practically unchanged during propagation.}
		\label{fig:properties}
	\end{figure}
	
	Caustic beams can be classified into three categories based on their mathematical derivation and physical origin. 
	The first category emerges from wave optics. By transforming the Helmholtz equation into cylindrical, elliptical, or parabolic cylindrical coordinates, we can obtain separable eigenmode solutions such as the Bessel, Mathieu, and Weber functions~\cite{Zannotti2020}, respectively. The transverse field distribution of these beams is intrinsically connected to the geometrical structure of the corresponding coordinate system employed in the solutions. 
	
	The second category originates from geometrical optics, where the radiation intensity is determined by the local density of passing rays. Singularities may arise at the location where adjacent rays are tangent to each other~\cite{liu2026}. 
	Strong EM wave intensity ridges then occur at these locations, which form caustic beams. 
	{These caustic beams frequently exhibit bending trajectories and retrieve their transverse structures after partial obstructions.} 
	Catastrophe theory is considered a unified framework to classify different singular structures, such as the fold, the cusp, the swallowtail, and the umbilic~\cite{BERRY1980257}.

	The third category directly originates from the generalized Snell's law~\cite{generalizedsnell}, which indicates that the {\it gradient} of the phase determines the local angle of departure (AoD) of the radiated EM wave.
	For example, for a linear aperture along the $x$-axis, the following differential relationship holds
	\begin{equation}
		\frac{\partial \phi(x)}{\partial x} = \frac{2\pi}{\lambda} \cos \theta, \label{eq:eikonal_x}
	\end{equation}
	where $\phi(x)$ is the aperture phase profile, $\theta$ denotes the AoD of the EM wave, and $\lambda$ is the wavelength. 
	By tailoring $\phi(x)$ so that the departing EM waves become tangent to a designated curve $y=f(x)$, one obtains a caustic beam whose main lobe follows that curved trajectory.

	The transition from conventional beams to caustic beams introduces three unique properties that are particularly useful for wireless communication scenarios:
	
	\begin{table*}[t]
		\centering
		\small
		\renewcommand{\arraystretch}{1.0}
		\begin{threeparttable}
			\caption{Properties of Representative Caustic Beams}
			\label{tab:beams}
			\begin{tabular}{@{}l l l l c c c@{}}
				\toprule
				\textbf{\makecell[c]{Beam\\ Family}} & 
				\textbf{\makecell[c]{Coordinate\\ System}} & 
				\textbf{\makecell[c]{Propagation\\ Trajectory}} & 
				\textbf{\makecell[c]{Transverse\\ Shape}} & 
				\textbf{\makecell[c]{Self-\\ Bending}} & 
				\textbf{\makecell[c]{Self-\\ Healing}} & 
				\textbf{\makecell[c]{Non-\\ Diffracting\tnote{$\dagger$}}} 
				\\ 
				\midrule
				
				\multicolumn{7}{c}{\textit{Helmholtz Eigenmode Solutions}} \\
				\midrule
				
				\textbf{Bessel} & 
				Cylindrical & 
				Rectilinear & 
				Concentric Rings & 
				$\bigcirc$ & 
				$\checkmark$ & 
				$\checkmark$ 
				\\ 
				
				\textbf{Mathieu} & 
				Elliptic & 
				Rectilinear & 
				Elliptical Arcs & 
				$\bigcirc$ & 
				$\checkmark$ & 
				$\checkmark$ 
				\\ 
				
				\textbf{Weber} & 
				Parabolic & 
				Rectilinear & 
				Parabolic Arcs & 
				$\bigcirc$ & 
				$\checkmark$ & 
				$\checkmark$ 
				\\ 
				
				\midrule
				\multicolumn{7}{c}{\textit{Catastrophe Theory-based Caustics~\cite{BERRY1980257}}} \\
				\midrule
				
				\textbf{Airy} & 
				Cartesian & 
				Parabolic & 
				Airy Function & 
				$\checkmark$ & 
				$\checkmark$ & 
				$\bigcirc$
				\\ 
				
				\textbf{Pearcey} & 
				Cartesian & 
				\makecell[l]{Auto-focusing Curve} & 
				Cusp & 
				$\checkmark$ & 
				$\checkmark$ & 
				$\bigcirc$
				\\ 
				\textbf{Swallowtail} & 
				Cartesian & 
				\makecell[l]{Complex Curve} & 
				Swallowtail-like & 
				$\checkmark$ & 
				$\checkmark$ & 
				$\bigcirc$
				\\ 
				\textbf{Umbilic} & 
				Cartesian & 
				\makecell[l]{Complex Curve} & 
				Multiple Cusps & 
				$\checkmark$ & 
				$\checkmark$ & 
				$\bigcirc$
				\\ 
				\midrule
				\multicolumn{7}{c}{\textit{Generalized Snell's Law-based Caustics~\cite{generalizedsnell} }} \\
				\midrule %
				
				\textbf{Eikonal} & 
				Cartesian & 
				\makecell[l]{Arbitrary $y = f(x)$} & 
				Locally Airy & 
				$\checkmark$ & 
				$\checkmark$ & 
				$\bigcirc$
				\\ 
				\bottomrule
			\end{tabular}
			
			\begin{tablenotes}
				\footnotesize
				\item[$\dagger$] The non-diffracting property takes effect for ideal beams with infinite aperture. For truncated practical beams, quasi-non-diffracting holds over a finite near-field region. $\bigcirc$ indicates that the property holds under certain circumstances.
			\end{tablenotes}
		\end{threeparttable}
	\end{table*}
	
	\begin{itemize}
		\item {\bf Self-Bending}: In a homogeneous and non-dispersive medium, the main intensity lobe of a monochromatic caustic beam can propagate along a designated curved trajectory, as shown in Fig.~\ref{fig:properties}(a). 
		While this may seem counter-intuitive, this phenomenon does not violate the rectilinear propagation principle of EM waves. 
		The main lobe of a beam arises from the coherent superposition of EM waves at specific locations, and the locations are not always fixed.
		For instance, in conventional far‑field beam steering, the main lobe moves proportionally with the propagation distance, resulting in a fixed‑angle beam; in near‑field beam focusing, the coherent superposition occurs only at the focal point, and the beam gain rapidly diverges away from it. 
		By carefully tailoring the amplitude and the phase profiles across the transmit array, we can design beams with a main intensity lobe that accelerates as it propagates. 
		This creates the spatial visual effect of a curved beam, and is the reason why such beams are called \textit{self‑accelerating} beams. 
		A typical example is the Airy beam, which can be realized either by applying an Airy function for the transmission amplitude profile~\cite{darsena2025airybeamsnearfieldcommunications} or by engineering a cubic phase~\cite{11267239}. 
		\item {\bf Self-Healing}: Caustic beams exhibit a remarkable self‑healing capability. 
		When the main lobe is partially obstructed (e.g., by a pedestrian or a drone) along the caustic trajectory, as illustrated in Fig.~\ref{fig:properties}(b), the transverse beam structure is temporarily disrupted. 
		Nevertheless, during subsequent propagation, the unobstructed portions continue to propagate and interfere with each other, reforming the main lobe and the original caustic trajectory. 
		{This is because caustic beams are engineered to realize prescribed interference patterns over the entire radiation region rather than at local points. 
		Consequently, partial obstructions mainly introduce local perturbations and do not destroy the global wavefront structure. 
		Self-healing is therefore a generic feature of caustic beams, indicating their potential for improved robustness against blockage, interference, and propagation interruptions in complex environments.}
		
		\item {\bf Near-Field Non-Diffracting}: In conventional near‑field beam focusing, the transverse field distribution of the beam changes during propagation. The focusing beam achieves its maximum gain only at the intended target location, and its transverse structure collapses quickly as the wave moves away from the target. 
		In contrast, caustic beams can preserve their transverse structure over an extended propagation distance by coherently organizing energy onto propagation-invariant wavefronts. 
		These wavefronts are often truncated approximations of propagation-invariant solutions to the Helmholtz equation, such as the Bessel, Mathieu, and Weber~\cite{Zannotti2020} functions, etc., thereby maintaining an approximately constant radiation profile within a localized region of space. An example of a Bessel beam is shown in Fig.~\ref{fig:properties}(c), where the beamwidth and the transverse structure remain practically {unchanged} along the transmission direction. 
		This property may alleviate the need for frequent beam training and estimation for mobile users and help reduce the signaling overhead challenge associated with conventional beamforming.
	\end{itemize}

	A taxonomy and properties of representative caustic beams are summarized in Table~\ref{tab:beams}.%

	\section{Applications in Next-Generation Wireless Communications}

	The unique propagation characteristics of caustic beams, namely self-bending, self-healing, and non-diffracting, open up a new design paradigm for 6G and beyond. In this section, 
	we discuss some potential applications of caustic beams in next-generation wireless communication systems, and provide results for physical layer security and reliable wireless coverage applications.

	\subsection{Resilient Wireless Connectivity}
	
	In next-generation wireless networks, where the mm-wave and THz bands will be utilized, communication links can be particularly fragile due to the propagation characteristics at very high frequencies. Simple obstructions of the LoS path often lead to link failure. 
	Caustic beams may offer a fundamental improvement in link reliability through their self-healing capability. When the main lobe encounters a partial obstruction, such as a pedestrian, the transverse structure of the caustic beam at the intended user farther along the propagation path remains largely intact. 
	This mechanism allows the communication link to survive dynamic scatterers that would otherwise block the connection, thereby enhancing the robustness of vehicular and industrial networks.

	\subsection{Reliable Wireless Coverage}
	Service stability characterizes a system's capability to sustain user quality of experience (QoE). In confined indoor scenarios with rich scattering and frequent blockages, the receive power can fluctuate with small position changes over time. Such instability can lead to dramatic QoE variations and may even trigger outages. 
	Caustic beams are able to achieve uniform coverage by enabling flexible synthesis of the desired radiation pattern according to the environment and user activities. Exploiting the self-bending property, caustic beams can be synthesized to uniformly cover the desired service region, rather than concentrating energy into a narrow mainlobe or leaving coverage to chance.

	\subsection{Robust Links for Robots/Embodied AI Agents}
	
	Embodied AI agents (e.g., autonomous mobile robots in factories, unmanned aerial vehicles, humanoids, etc.) require ultra-reliable and low-latency links while suffering from frequent blockage, pose variations, and changing scattering environments. These requirements become particularly stringent in high-frequency bands where non-LoS links are more difficult to establish.%
	
	Caustic beams may provide a unique opportunity to enhance link robustness by leveraging their unique propagation characteristics. Specifically, the non-diffracting behavior ensures a relatively constant energy distribution along the moving trajectory, while the self-bending and self-healing properties improve the link robustness of the mobile embodied agents.
	
	\subsection{Physical Layer Security} %

	In wireless systems, an eavesdropper may want to intercept the signal sent by a legitimate transmitter to an intended user. 
	The secrecy rate, i.e., the rate advantage of the legitimate link over the eavesdropping link, is a key measure of confidentiality. 
	Practically, as the channel state information of the eavesdropper cannot be precisely acquired, uncertainty regarding the eavesdropper's location is essentially unavoidable. 
	As a result, eavesdropping may occur {in a neighborhood around the estimated position of the eavesdropper}. 
	To ensure secure communication, the wireless system should guarantee a certain worst-case secrecy rate for any possible eavesdropper location within this region.

\begin{figure}[t]
		\centering
		\includegraphics[width=0.5\textwidth]{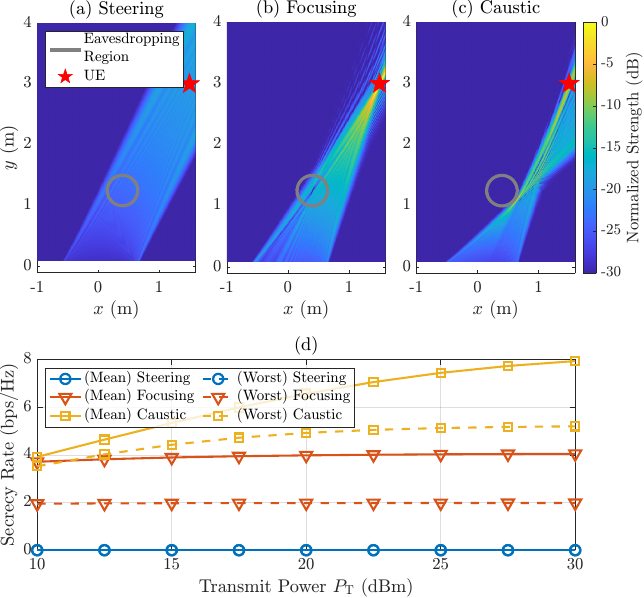}%
		\caption{The radiation patterns of (a) a conventional steering beam, (b) a secure focusing beam, and (c) the proposed caustic beam, and (d) the secrecy performance versus the transmit power for the designs shown in (a)-(c): A $256$-element transmit array with half-wavelength spacing operates at $30$ GHz and is placed along the
		$x$-axis. The UE is located at $(1.5,3)$ m, while the gray circle denotes the eavesdropper uncertainty
		region centered around its estimated location $(0.4,1.2)$ m, with a localization uncertainty radius of $0.25$ m.
		}
		\label{fig:secure}
	\end{figure}

	Caustic beams provide a geometrically intuitive and practically appealing mechanism to enhance the link security under localization uncertainty~\cite{liu2026}. 
	Thanks to their self‑bending property, caustic beams can be synthesized to follow prescribed curved trajectories. 
	The beam can then be shaped to deliberately bypass the whole potential eavesdropper region, thereby ensuring a high worst-case secrecy rate, even when the eavesdropper is at the most unfavorable position within the uncertainty region, as shown in Fig.~\ref{fig:secure}.

	\subsection{Stable Near-Field Service with User Mobility}

	The non-diffracting property of various caustic beams yields an extended depth of the radiation field in the near-field region. Unlike conventional near-field focusing beams, which suffer from rapid energy attenuation outside the focal point, non-diffracting beams sustain a stable transverse energy density over a significantly longer propagation distance on the designated trajectory. For mobile near-field users, this property ensures seamless and stable service, which also alleviates the massive signaling overhead associated with frequent beam training and tracking at the base station~\cite{wang2026}.
	Besides, non-diffracting beams maintain a highly stable beamwidth along their propagation trajectories, indicating a consistent, distance-independent spatial resolution within the near-field region. This unlocks significant potential for various applications, including near-field environment sensing, industrial non-destructive testing, and high-resolution microwave imaging.

\begin{figure}[t]
		\centering
		\includegraphics[width=0.5\textwidth]{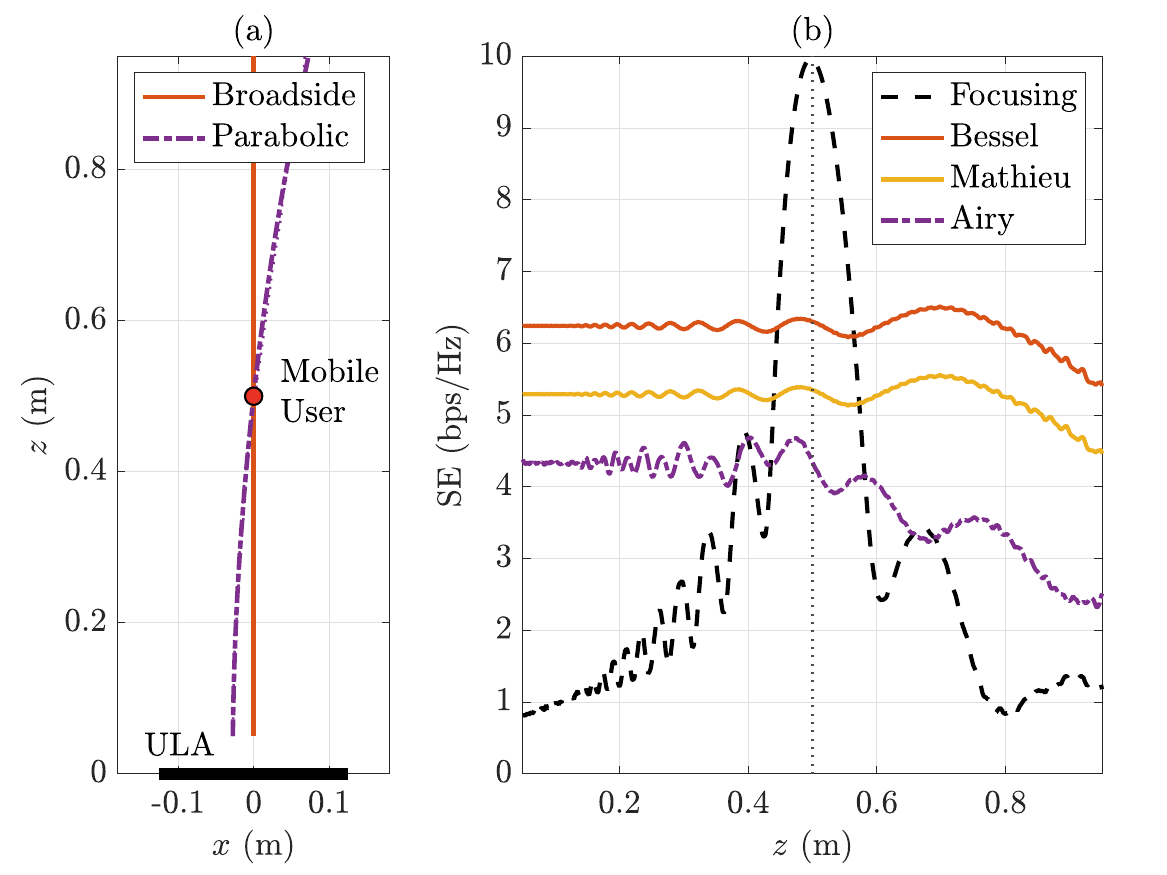}%
		\caption{(a) Geometric setting of the near-field communication system and trajectories of the mobile user. (b) SE performance versus user's $z$-position along the corresponding trajectory.}
		\label{fig:nondiff}
	\end{figure}

	To illustrate the benefits of non-diffracting beams in the near field, {we consider the geometric setting shown in Fig.~\ref{fig:nondiff}(a). A transmit array with a $0.25$~m aperture operates at $100$~GHz and serves a mobile user in the near-field region. Two typical trajectories, i.e., a broadside (for focusing, Bessel, and Mathieu beams) and a parabolic (for Airy beam) trajectory, are considered for a mobile user. 
	As shown in Fig.~\ref{fig:nondiff}(b), beam focusing achieves the highest spectral efficiency (SE) at the focusing point $z_0 = 0.5$ m. However, 
	the truncated non-diffracting Bessel and Mathieu beams outperform the focusing scheme when the user's movement distance $\Delta z = \vert z-z_0\vert$ exceeds $0.05$ m along the broadside
	direction.} 
	{To further evaluate the performance under a non-rectilinear service trajectory, we also test the Airy beam along its parabolic self-bending trajectory, which exhibits a similar robustness trend according to Fig.~\ref{fig:nondiff}.} 
	As $\Delta z$ further increases, the SE performance of the focusing scheme keeps decreasing, while all considered caustic beams maintain a relatively stable SE performance across the entire observation window of $0.9$ m. 
	Hence, in practice, to maintain a stable SE performance along a trajectory, the focusing scheme needs to switch the beam focal point frequently, which incurs a high signaling overhead. This overhead can be significantly reduced when employing non-diffracting caustic beams.

	\section{Enabling Architectures}
	
	Transferring the theoretical advantages of caustic beams to practical wireless systems requires 
	hardware capable of shaping the transmitted wavefront with sufficient accuracy over a physically large aperture. 
	However, practical hardware architectures are subject to different constraints such as finite aperture size, discrete elements, quantized amplitude-phase control, etc. Therefore, caustic beam synthesis involves a trade-off between wavefront fidelity and implementation complexity. 
	In this section, we categorize these enabling architectures based on their per-element manipulation degrees of freedom.

	\subsection{Full Amplitude and Phase Control}
	
\begin{figure}[t]
		\centering
		\includegraphics[width=0.5\textwidth]{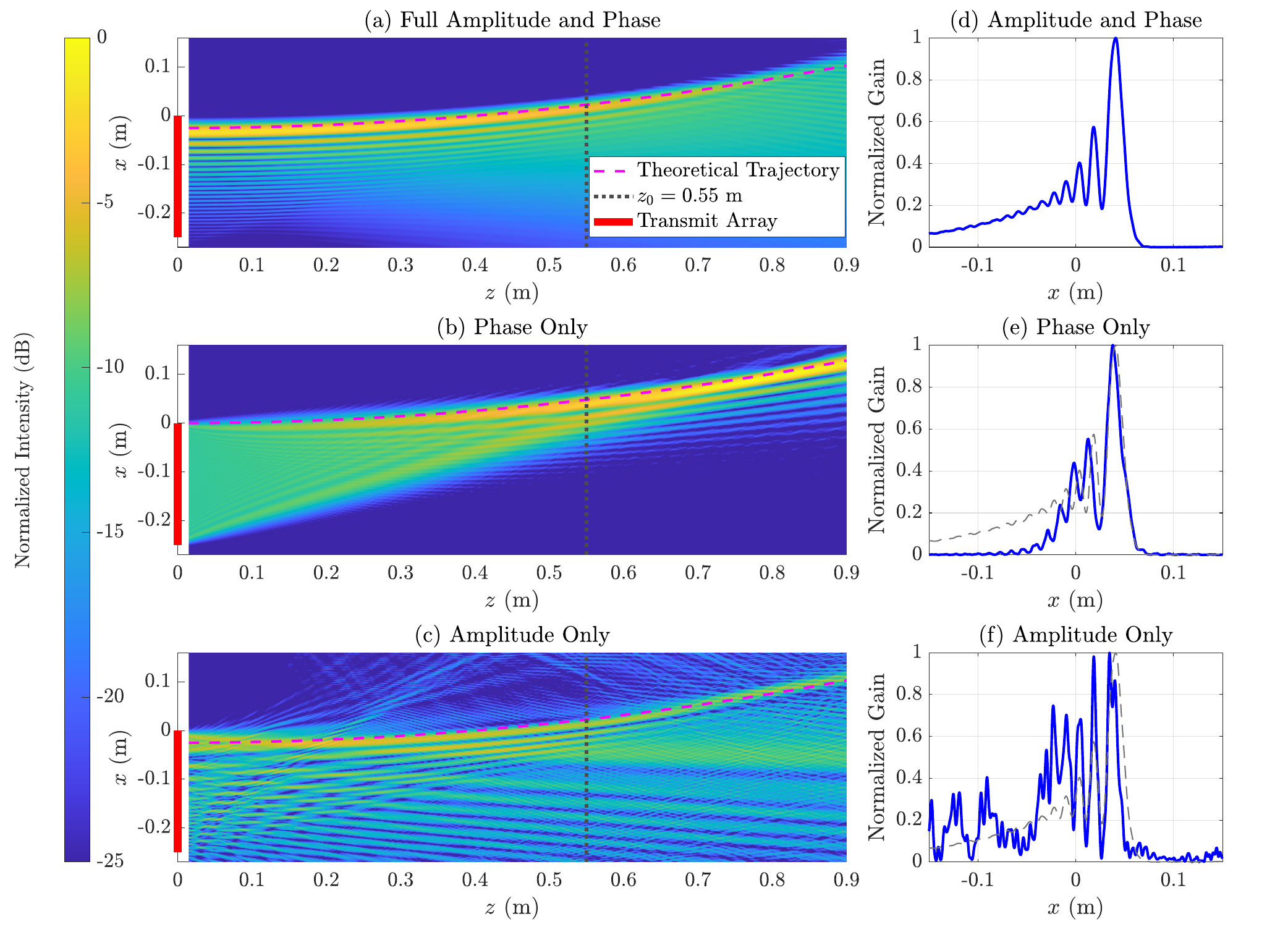}%
		\caption{Caustic beam synthesized by representative hardware architectures: (a) Fully digital phased array, (b) phase plate, and (c) on-off amplitude only metasurface. 
		(d)-(f) Corresponding transverse patterns at $z_0 = 0.55$ m. All architectures target the same parabolic trajectory $x = z^2/6.25$ at a carrier frequency of $60$ GHz with the aperture of $0.25$ m marked in red.
		}
		\label{fig:hardware}
	\end{figure}

	The most powerful architecture for synthesizing caustic beams is the fully digital phased array, where each radiating element is driven by an independent RF chain to manipulate both amplitude and phase of the transmitted EM wave. 
	By precisely controlling each radiating element, phased array systems can directly synthesize non-diffracting solutions of the Helmholtz equation or any arbitrary excitation function required for caustic beams, as listed in Table~\ref{tab:beams}. 
	This flexibility also allows a single aperture to superpose multiple caustic beams with distinct trajectories for multi-user service, and to re-synthesize the beam on a slot-by-slot basis as the environment and user location change~\cite{10791450}.
	As shown in Figs.~\ref{fig:hardware}(a) and (d), the fully digital architecture can synthesize caustic beams with both fine trajectory and transverse structure, which can be a reference benchmark for the other two considered architectures.

	While phased arrays offer the highest manipulation degree of freedom, real-time reconfigurability, and multi-caustic beam synthesis, such flexibility comes at the cost of significant hardware complexity and power consumption. These overheads become particularly prohibitive for caustic beam applications, as a large physical aperture is inherently required to maintain the unique propagation characteristics of caustic beams.
	To amortize such overheads, hybrid analog-digital architectures are an attractive alternative requiring a moderate number of RF chains. 
	Besides, for wideband operations at mm-wave and THz frequencies, true-time-delay (TTD) elements can also be adopted in the feed network to mitigate the beam squint effect.

	\subsection{Synthesizing via Phase}

	To further mitigate implementation costs and enhance feasibility, current research works frequently adopt phase-only manipulation structures~\cite{10964575,guerboukha_curving_2024}. As the amplitude is fixed, these structures typically synthesize caustic beams by implementing carefully designed phase profiles derived from~\eqref{eq:eikonal_x} or numerical optimization methods. 
	
	Phase plates fabricated by 3D printing~\cite{guerboukha_curving_2024} provide a cost-effective and precise way to synthesize static caustic beams. They locally modify the phase of an incident beam by tailoring the thickness of the plate.
	Moreover, reconfigurable intelligent surfaces (RIS) employing positive-intrinsic-negative (PIN) diodes or varactors provide tunable yet discrete phase profiles for dynamic scenarios, while liquid crystal metasurfaces~\cite{liquid_crystal} extend continuous phase control into even higher frequency bands, i.e., THz, at the cost of slower reconfiguration.

	A caustic beam synthesized by a phase plate is illustrated in Fig.~\ref{fig:hardware}(b), with the corresponding transverse structure given in Fig.~\ref{fig:hardware}(e). Although arbitrary trajectories can be achieved, phase-only architectures are generally less flexible than fully digital phased arrays in controlling the radiation pattern. Specifically, these architectures can only synthesize the generalized Snell's law-based caustic beams in Table~\ref{tab:beams}, and lack the capability to manipulate the power allocation alongside the designated caustic trajectory.
	
	\subsection{Synthesizing via Amplitude}
	
	Another approach is to synthesize caustic beams with amplitude-only manipulation. Typical architectures, including leaky-wave antennas, holographic surfaces, and on-off switch-controlled surfaces, reduce hardware complexity and power consumption by eliminating RF chains and even phase shifters. Specifically, leaky-wave antennas radiate a guided reference wave via impedance modulation, while holographic surfaces reconstruct the target beam by exciting the interference pattern, i.e., hologram, with the same reference wave~\cite{10845870}. On-off switch-based surfaces use binary amplitude coding to achieve a desired radiation pattern. 
	
	Nevertheless, achieving high-fidelity caustic beams with amplitude-only control remains challenging. The lack of phase control narrows the realizable caustic family, since most of the beams listed in Table~\ref{tab:beams} require carefully designed phase profiles. From the algorithm perspective, the reduced manipulation degrees of freedom require more sophisticated iterative and combinatorial optimization to suppress synthesis errors. As illustrated in Fig.~\ref{fig:hardware}, although the overall caustic trajectory is still recognizable, the transverse pattern exhibits significant distortion.

	\section{Challenges and Future Directions}
	
	While caustic beamforming opens a new design paradigm to shape the propagation trajectory and the spatial energy distribution of EM waves, several technical challenges still have to be overcome before its practical adoption in 6G wireless systems will be possible. In this section, we discuss key issues and promising future research directions.%
	
	\subsection{Diffraction-Aware Channel Modeling and Acquisition}
	
	The outstanding properties of caustic beams discussed in Section~\ref{sec:iii} are caused by the diffraction effect of EM waves. However, most channel models adopted in current wireless studies are built upon geometric channel models and explicitly discard the contribution of diffraction. Therefore, conventional channel models may severely underestimate the received field in the shadowed region, since the diffraction effect can still deliver non-negligible power to the receiver even when the direct LoS path is obstructed. 
	An important research direction is the development of diffraction-aware channel models in a matrix-friendly form~\cite{11267239}.

	{Accurate caustic beamforming requires channel models that capture diffraction paths, blockage geometry, and dominant scatterers~\cite{11267239}. Therefore, environmental sensing is essential for diffraction-aware channel acquisition and adaptive beam synthesis. ISAC techniques can be utilized to estimate the locations of scatterers, obstacles, and intended users, which provides the geometric information needed for diffraction-aware channel construction. Thus, the joint design of caustic sensing and beamforming is an open problem worth pursuing.}

	\subsection{Universal Caustic Beamforming Algorithms}
	
	The flexible locations of obstacles and users in real environments pose fundamental challenges for the synthesis of arbitrary caustic beams. Users may occupy different positions in 3D space, while obstacles can create volumetric blockages. 
	Most caustic beams listed in Table~\ref{tab:beams} follow fixed patterns. 
	{Although the Eikonal-based design can support arbitrary caustic trajectories, its 3D extension is particularly challenging because the prescribed 3D caustic surfaces must be mapped to the phase gradients on the aperture, which are then integrated to obtain the transmit phase. }

	A promising approach to design complex caustic beams is to utilize AI tools to learn the mapping from 3D environment embeddings to the caustic amplitude-phase profile~\cite{Chen2025}. 
	Such models can be trained with physics-informed features and parameters to predict near-optimal amplitude-phase profiles with negligible inference overhead. Nevertheless, algorithms that achieve low complexity and high robustness still require further investigation.

	\subsection{Hardware Impairments}
	In practical wireless systems, hardware impairments are inevitable. Caustic beamforming relies on precise control of EM wave diffraction and is therefore particularly sensitive to hardware impairments. The most fundamental impairment arises from the mismatch between the discrete antenna configuration on practical arrays and the continuous amplitude-phase excitation profiles required to synthesize caustic beams. Similarly, conventional phase shifters usually suffer from limited quantization resolution, while the analytically derived excitation functions for caustic beams are usually continuous in phase.
	As operating frequencies increase to meet the needs of higher bandwidth, phase shifter-based beamforming architectures become vulnerable to the beam squint effect, which degrades system performance by generating undesired beam components. 
	Furthermore, phase plates may exhibit noticeable amplitude-phase coupling, making it difficult to independently control the aperture amplitude and phase, which may distort the intended caustic trajectory.

	Mitigating these hardware impairments requires future research on both architectures and algorithms. Emerging architectures such as continuous-aperture arrays (CAPAs) and TTD networks can better approximate the continuous excitation profiles, while advanced algorithms for impairment calibration and compensation are also desired for future deployments.
	
	\section{Conclusions}
	In this article, we introduced the emerging caustic beam technique that moves wireless beamforming beyond conventional far-field steering and near-field focusing. We further presented the unique characteristics of caustic beams, and discussed their applications as well as future research challenges. In summary, the self-bending, self-healing, and non-diffracting capabilities of caustic beams allow EM energy to be routed along curved trajectories, regenerated after partial blockage, and sustained over an extended depth of field range.  
	However, exploiting such sophisticated beams 
	in practical applications requires significant efforts in diffraction-aware channel modeling, universal caustic beamforming algorithm design, as well as calibration and compensation of hardware impairments. 
	Caustic beams have the potential to play a crucial role in the ongoing migration of wireless systems to the mm-wave and THz frequency bands and to complement traditional beam steering and focusing in future wireless networks.

	\bibliographystyle{IEEEtran}
	\bibliography{IEEEabrv,references}
	\end{document}